\documentclass[aps,prl,twocolumn,superscriptaddress]{revtex4}

\usepackage{graphicx}
\usepackage{amssymb}
\usepackage{xcolor}

\begin{document}

\title{Ordering Nanoparticles with Polymer Brushes}

\author{Shengfeng Cheng}
\email{chengsf@vt.edu}
\affiliation{Department of Physics, Center for Soft Matter and Biological Physics, and Macromolecules Innovation Institute, Virginia Polytechnic Institute and State University, Blacksburg, Virginia 24061, USA}
\author{Mark J. Stevens}
\affiliation{Sandia National Laboratories, Albuquerque, NM 87185, USA}
\author{Gary S. Grest}
\affiliation{Sandia National Laboratories, Albuquerque, NM 87185, USA}

\date{\today}

\begin{abstract}
Ordering nanoparticles into a desired super-structure is often crucial for their technological applications. We use molecular dynamics simulations to study the assembly of nanoparticles in a polymer brush randomly grafted to a planar surface as the solvent evaporates. Initially, the nanoparticles are dispersed in a solvent that wets the polymer brush. After the solvent evaporates, the nanoparticles are either inside the brush or adsorbed at the surface of the brush, depending on the strength of the nanoparticle-polymer interaction. For strong nanoparticle-polymer interactions, a 2-dimensional ordered array is only formed when the brush density is finely tuned to accommodate a single layer of nanoparticles. When the brush density is higher or lower than this optimal value, the distribution of nanoparticles shows large fluctuations in space and the packing order diminishes. For weak nanoparticle-polymer interactions, the nanoparticles order into a hexagonal array on top of the polymer brush as long as the grafting density is high enough to yield a dense brush. An interesting healing effect is observed for a low-grafting-density polymer brush that can become more uniform in the presence of weakly adsorbed nanoparticles.
\end{abstract}

\maketitle

\section{Introduction}
When polymer chains are end-tethered to a surface, a polymer brush is formed, which provides a convenient means to alter the properties of the surface.\cite{mansky97} Introducing nanoparticles (NPs) into the brush gives rise to a wide range of new hybrid materials.\cite{ferhan16,nie16} Such brush-NP systems have broad applications in fields including sensors, information storage devices, medical diagnostic tools, and catalysts, depending on the combination of the NPs and the polymer brush.\cite{christau15} The brush-NP hybrids also serve as a model system to study drug delivery.\cite{mullner16} Furthermore, if the polymer brush is responsive to some perturbations or signals, then the hybrid materials can be used as active components to make various ``smart'' systems.\cite{tokarev12}

For a given brush-NP composite, the state of the brush and the distribution of NPs in the brush are critical factors that determine its prospect for particular applications. These considerations have motivated many experimental studies on controlling the NP distribution and organization in a polymer brush so that certain functions can be realized.\cite{christau15} The hybrid structure is determined collectively by the strength of enthalpic NP-polymer interaction,\cite{diamanti09} the brush characteristics including grafting density,\cite{ferhan12} thickness,\cite{liu02} pattern,\cite{onses12,onses13,steinbach13} composition,\cite{oren09} and possible gradients of these properties,\cite{bhat03,bhat04,bhat06} and the solvent being used.\cite{kesal16}

Various theoretical and computational techniques have been employed to study the interactions between NPs and a polymer brush, including the self-consistent field theory (SCFT),\cite{subramanian96,solis96,steels00,kim02,kim06,kim08,chen05,devos09,devos10,halperin11,egorov12,lian15} Monte Carlo simulations,\cite{milchev08,ermilov10,chen11} molecular dynamics (MD) simulations,\cite{yaneva09,merlitz12,yigit17} dissipative particle dynamics (DPD),\cite{chengjianli15} and Brownian dynamics simulations. \cite{zhang13PhysicaA} These and related methods have also been used to examine various strategies of controlling the distribution and organization of NPs in a polymer brush.\cite{cao07,guskova09,milchev10,opferman12,opferman13,curk13,curk14,zhang13SoftMatter, zhang14JPSB,hua16} For example, Cao and Wu explored the idea of using block copolymer brushes to organize NPs into a 3-dimensional superstructure or a 2-dimensional (2D) layer via a density functional theory.\cite{cao07} Guskova \textit{et al.} observed the cluster formation of polymer-insoluble NPs at the surface of a brush with DPD simulations.\cite{guskova09}  Zhang \textit{et al.} showed via MD simulations that a polymer brush can be used to induce phase separation and crystallization of a binary mixture of NPs embedded in the brush by tuning the size ratio of NPs and the NP-polymer interactions.\cite{zhang13SoftMatter, zhang14JPSB} Hua \textit{et al.} later found that for a similar system the phase separation can also be induced by compressing the NP-containing brush.\cite{hua16}

Despite extensive studies on NP-brush hybrid materials, the effects of solvent evaporation in these systems have been barely explored, even though solvent evaporation is frequently employed in experiments and manufacturing to make polymer-NP composites. Many studies have shown that the evaporation rate plays a critical role in the drying of polymer films and paint.\cite{composto90,strawnhecker01,koombhongse01,luo05,erkselius08,zhang12,kooij15} Jouault \textit{et al.} found that using different casting solvents the same NPs will either aggregate or disperse in the same polymer matrix.\cite{jouault14b} Cheng and Grest showed via MD simulations that the dispersion of NPs in a polymer film is strongly influenced by the evaporation process of the solvent.\cite{cheng16} Kumar \textit{et al.} emphasized the role of solvent evaporation as an outstanding theoretical question in polymer-NP hybrids in a recent review.\cite{kumar2017JCP} Solvent evaporation has also been frequently employed to assemble NPs into various patterns and superstructures.\cite{brinker99,hamon14,josten17,ryu17} It is thus interesting to study the role of solvent evaporation on the structure of brush-NP hybrids.

\begin{figure}[ht]
\centering
\includegraphics[width=3.25in]{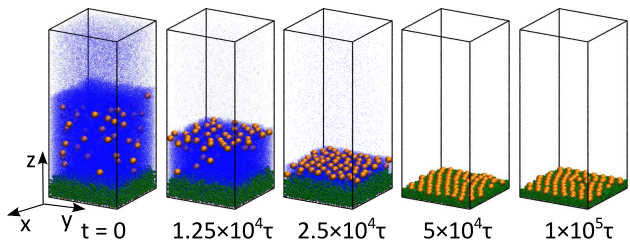}
\caption{Snapshots at various stages during solvent evaporation for the system with $\phi=0.08\sigma^{-2}$, $C=0.65$, and $A_{\rm np} = 100\epsilon$ (see Model and Methodology for the definition of $\phi$, $C$, and $A_{\rm np}$). Color codes: solvent (blue), polymer brush (green), and NPs (orange).}
\label{evolution}
\end{figure}

Here we report on MD simulations of how NPs, initially dispersed in an explicit solvent wetting a polymer brush, are distributed in the brush as the solvent is evaporated. We are particularly interested in determining the condition for a monolayer of NPs to form a well ordered structure (i.e., a hexagonal lattice). One example is shown in Fig.~\ref{evolution}. Our results reveal that depending on brush-NP interactions, various strategies can be used to assemble a well-ordered 2D lattice of NPs, either in or on top of the brush, after solvent evaporation. For strong brush-NP interactions, a polymer brush with a thickness suitable to adsorb a single layer of NPs is needed to yield an ordered 2D array of NPs embedded in the brush. For weak brush-NP interactions, NPs form an ordered 2D array on top of the brush as long as the grafting density is high enough (Fig.~\ref{evolution}). We also find an interesting healing effect where a polymer brush with a low grafting density can cover the grafting plane in a more uniform manner in the presence of weakly adsorbed NPs.

\section{Model and Methodology}
We model the solvent as a Lennard-Jones (LJ) liquid and the polymer as linear chains of 100 beads connected by permanent bonds. All solvent and polymer beads have a mass $m$ and interact through a standard LJ 12-6 pairwise potential, $U_{\rm LJ}(r)=4\epsilon [(\sigma/r)^{12}-(\sigma/r)^6 -(\sigma/r_c)^{12}+(\sigma/r_c)^6 ]$, where $r$ is the distance between the centers of two beads, $\epsilon$ the strength of interaction, and $\sigma$ the size of beads. The cut off distance is $r_c=3.0\sigma$ for all non-bonded pairs and $r_c=2^{1/6}\sigma$ for bonded pairs of neighboring polymer beads on a chain. The latter are connected by a bond given by the finitely extensible nonlinear elastic potential, $U_{B}(r) = -\frac{1}{2} K R_0^2 {\rm ln}[ 1-( r/R_0 )^2 ]$, where $r$ is the separation distance, $R_0=1.5\sigma$, and $K=30\epsilon/\sigma^2$.\cite{kremer90}

\begin{figure}[ht]
\centering
\includegraphics[width=3in]{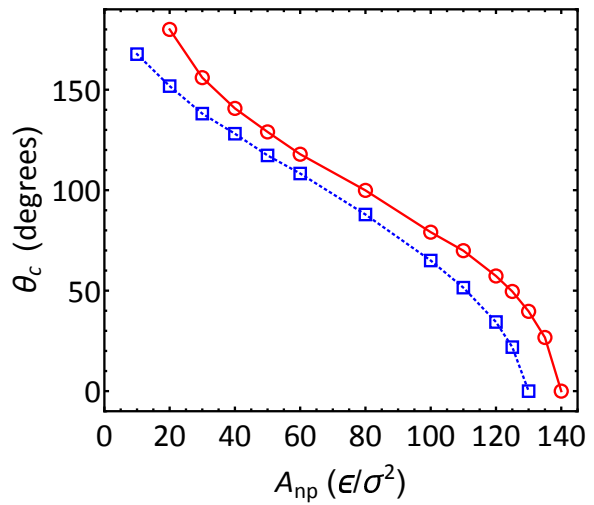}
\caption{Contact angle $\theta_c$ vs. $A_{\rm np}$ for $D=10\sigma$ (circles) and $\infty$ (squares). Lines are guides to the eye.}
\label{angle}
\end{figure}

The NPs are modeled as spheres of a diameter $d=10\sigma$. Each NP is treated as a uniform distribution of LJ point masses $m$ at a density $1.0m/\sigma^3$, resulting in a mass $M=523.6m$ for the NPs. The NP-NP interaction is described by an integrated form of the LJ 12-6 potential, and its strength is characterized by a Hamaker constant $A_{\rm nn} = 39.48\epsilon$.\cite{Everaers03} To avoid flocculation, the interaction between NPs is chosen to be purely repulsive with the cut off set at $10.438\sigma$.\cite{veld09,grest11} The interaction between a NP and a solvent or polymer bead is determined by a similar integrated potential with Hamaker constants $A_{\rm ns}$ and $A_{\rm np}$, respectively; both interactions are truncated at $9\sigma$. We set $A_{\rm ns}=100\epsilon$, which is sufficiently strong to disperse the NPs in the pure solvent without any polymer chains.\cite{cheng12JCP} We compare two systems with different NP-polymer interaction strengths: $A_{\rm np}=100\epsilon$ and $A_{\rm np}=200\epsilon$. The strength determines the contact angle $\theta_c$ of a NP floating at the surface of a polymer melt that consists of the 100-bead chains modeled here. The results for $\theta_c$ vs. $A_{\rm np}$ are included in Fig.~\ref{angle}. For $A_{\rm np}=100\epsilon$, $\theta_c = 79^\circ$, which indicates that the NPs are only partially adsorbed at the surface of the melt; while for $A_{\rm np}=200\epsilon$, $\theta_c = 0$, which indicates that the NPs initially placed at the surface of the melt eventually diffuse into the bulk of the melt. In Fig.~\ref{angle} we also show the contact angle of a drop formed by 2,346 chains of length 100 on a flat surface, with the same integrated potential for the polymer-surface interaction as that for the polymer-NP interaction extrapolated to $D=\infty$, as a function of $A_{\rm np}$. The data clearly show the effect on $\theta_c$ from the curvature of the NP surface; the smaller the NP, the higher the value of $\theta_c$ at a given $A_{\rm np}$ as long as $0<\theta_c<\pi$.

The model adopted here relies on using LJ potentials and their integrated forms to describe interactions between particles including polymer beads and is thus appropriate for systems dominated by van der Waals interactions. In many experimental systems, NPs and polymers may be charged in a polar solvent such as water and electrostatic interactions may play a major role in controlling the NP-polymer interactions.\cite{bhat03,bhat04,bhat06} In such systems, the NP-polymer interactions may be tuned by varying salt concentrations and exploiting the screening effect of mobile ions on electrostatic interactions. Such situations are of great interest for future studies. Although limited, studies using nonaqueous NP solutions to make NP-brush hybrids were also reported.\cite{oren09,snaith05}

The simulation cell is a rectangular box of dimensions $L_x \times L_y \times L_z$ with $L_x=L_y=100\sigma$. The liquid/vapor interface is parallel to the $x$-$y$ plane, in which periodic boundary conditions are imposed. Polymer chains are randomly end-grafted to the $x$-$y$ plane at $z=0.5\sigma$ with the minimum separation between the tether points set as $r_{tp}$, yielding a planar brush with a grafting density $\phi\equiv N_c/(L_x L_y)$ where $N_c$ is the number of chains. We vary $N_c$ from 100 to 2400; $\phi$ thus changes from $0.01\sigma^{-2}$ to $0.24\sigma^{-2}$. If we take $\sigma \simeq 0.5~{\rm nm}$, then the range of $\phi$ is roughly from $0.04~{\rm nm}^{-2}$ to $0.96~{\rm nm}^{-2}$. In all the simulations, the tether points are fixed and we set $r_{tp}=2\sigma$ except for $N_c=2400$ where $r_{tp}=1.6\sigma$. In the $z$ direction, the system is confined between two flat walls at $z=0$ and $z=L_z$, respectively. The top wall is far above the liquid/vapor interface below which the solvent, NPs, and polymer brush are located. The particle-wall interactions are represented with a LJ 9-3 potential $U_{\rm W}(h)=\epsilon_W [(2/15)(D/h)^{9}-(D/h)^3 -(2/15)(D/h_c)^{9}+(D/h_c)^3 ]$, where $\epsilon_W$ is the interaction strength, $D$ the characteristic length, $h$ the separation between the center of a particle and the wall, and $h_c$ the cut off separation. For all particles, $\epsilon_W=2.0\epsilon$. For the solvent and polymer beads, $D=1.0\sigma$ while for the NPs, $D=5.0\sigma$. We make all the walls repulsive for particles by setting $h_c=0.8583D$ except for the solvent beads at the lower wall, for which $h_c=3.0\sigma$ so that in the initial state the solvent does not dewet the lower wall and the brush.

The systems start in a state where $N$ NPs are randomly dispersed in the solvent that wets the polymer brush and is in equilibrium with the vapor phase of the solvent beads. We define the coverage of the NPs in the $x$-$y$ plane as $C\equiv N\sqrt{3}d^2/(2L_x L_y)$, in which case $C=1$ corresponds a 2D hexagonal close packed layer of NPs in contact that fully covers the $x$-$y$ plane. For the systems studied here, a full coverage with a close packed monolayer requires about 115 NPs. Since the NPs disperse in structures that are not closed packed, we focus primarily on coverages with $20 \le N \le 100$, which correspond to $0.17 \le C \le 0.87$ and can yield full monolayers. The number of solvent beads is varied based on $N_c$ and $N$ to ensure that all the NPs are initially dispersed in the brush-solvent mixture with a volume fraction lower than $10\%$. The approximate location along the $z$-direction for the liquid/vapor interface is estimated and the value of $L_z$ is chosen to give a vapor of thickness of at least $150\sigma$. The systems are then equilibrated for at least $2\times 10^4\tau$, with $\tau \equiv\sqrt{m\sigma^2/\epsilon}$. During equilibration, a vapor phase of the solvent beads is established, a liquid/vapor interface is formed, and the densities of solvent beads in the liquid and vapor phase reach their equilibrium values, $0.62 m/\sigma^{3}$ and $0.055 m/\sigma^{3}$, respectively.

We performed all the simulations using the Large-scale Atomic/Molecular Massively Parallel Simulator (LAMMPS).\cite{plimpton95,lammps} The equations of motion are integrated using a velocity-Verlet algorithm with a time step $\delta t =0.005\tau$. During the equilibration, the temperature $T$ is held at $1.0\epsilon/k_{\rm B}$ by weakly coupling all beads to a Langevin thermostat with a damping constant $0.1\tau^{-1}$. Once the liquid/vapor interface is equilibrated, the Langevin thermostat is removed except for solvent and polymer beads within $5\sigma$ of the lower wall.\cite{cheng11}

When the evaporation process of the solvent is initiated, the upper wall is moved to $L_z + 20\sigma$ with the region $[L_z, L_z + 20 \sigma]$ designated as the deletion zone. Unless noted, every $0.5\tau$ all the solvent beads in the deletion zone are removed, effectively mimicking the situation that the system is in contact with a vacuum. The evaporation rate which is defined as the number of beads removed every $\tau$ is high initially ($\sim$ $200/\tau$) and decreases with time quickly, approaching a plateau value ($\sim$ $20/\tau$) before diminishing to almost 0 in the late stage of evaporation when most of the solvent is evaporated.\cite{cheng11} We also conducted simulations with fixed, lower initial evaporation rates ($20/\tau$ or $10/\tau$) and found that the structure of the final brush-NP hybrid after solvent evaporation is insensitive to the evaporation rate, at least for the range of evaporation rates accessible to our MD simulations.

\section{Results and Discussion}

\begin{figure*}[ht]
\centering
\includegraphics[width=4.5in]{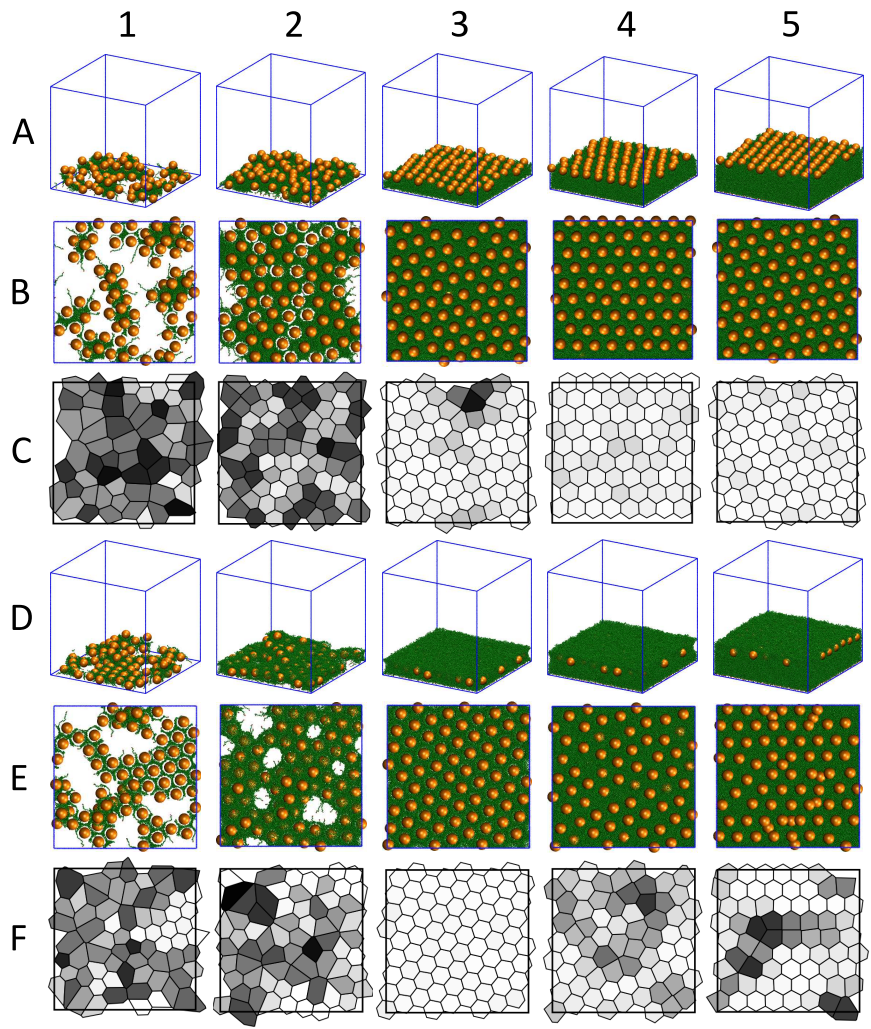}
\caption{Organization of NPs in brushes at various grafting densities after almost all the solvent has evaporated. For all the systems shown, $C=0.65$. For the top three rows (A-C), $A_{\rm np} = 100\epsilon$ while for the bottom three rows (D-F), $A_{\rm np} = 200\epsilon$. (A) and (D) side view; (B) and (E) top view; (C) and (F) Voronoi construction with gray scale based on the value of the Nelson-Halperin order parameter [see Eq.~(\ref{NHEq})] for the NP that each cell encloses. The value of $\phi$ is 0.01, 0.04, 0.08, 0.16, and 0.24$\sigma^{-2}$ for columns 1-5, respectively. In (3E)-(5E), to better show the embedded NPs the polymer beads covering them are not included. The color codes for rows (A), (B), (D), and (E) are the same as in Fig.~\ref{evolution}.}
\label{assembly}
\end{figure*}

Figure~\ref{evolution} illustrates an exemplary process of NP ordering at the surface of a polymer brush during solvent evaporation for $A_{\rm np} = 100\epsilon$. Initially, the NPs are randomly distributed in the solvent, which wets the brush. As the solvent evaporates quickly, the NPs start to accumulate at the receding liquid/vapor interface, because the diffusion of the NPs is much slower than the motion of the interface. Depending on the NP concentration, one or more layers of NPs are formed. For the right NP concentration and a NP-brush interaction leading to $\theta_c > 0$ and favoring phase separation such as in Fig.~\ref{evolution}, one NP layer is formed. In the final state, the NPs are adsorbed at the brush surface and form a 2D triangular lattice with defects.

Simulations such as the one in Fig.~\ref{evolution} are carried out for different combinations of the grafting density $\phi$, the NP coverage $C$, and the strength of the NP-polymer interaction set by $A_{\rm np}$. In Fig.~\ref{assembly} we focus on the effect of varying $\phi$ from 0.01$\sigma^{-2}$ to 0.24$\sigma^{-2}$, and contrast the cases with $A_{\rm np} = 100\epsilon$ ($\theta_c = 79^\circ$) to those with $A_{\rm np} = 200\epsilon$ ($\theta_c = 0$) at a fixed NP coverage $C=0.65$. Embedding a NP in the brush forces the polymer chains around the NP to displace and stretch, which leads to an entropy penalty. For $A_{\rm np} = 100\epsilon$, the enthalpic gain from embedding the NP is not sufficient to compensate for the entropy loss. As a result, in the equilibrium state before evaporation all the NPs are dispersed in the solvent, and the NPs are adsorbed at the surface of the polymer brush after the solvent is evaporated. When the grafting density is low ($\phi = 0.01\sigma^{-2}$), the brush forms islands after drying and the NPs straddle the surface of those islands, as shown in Figs.~\ref{assembly}(1A-1C). As the grafting density is increased to 0.04$\sigma^{-2}$, the brush starts to cover more of the substrate though holes still occur on the surface. The NP distribution after solvent evaporation follows the brush coverage, as shown in Figs.~\ref{assembly}(2A-2C). When the grafting density is further increased to 0.08$\sigma^{-2}$, the brush fully covers the substrate after solvent evaporation and forms a layer with a uniform thickness. After drying the NPs remain on top of the brush and form a lattice in which hexagonal order is easy to identify, as shown in Figs.~\ref{assembly}(3A-3C). Defects still occur and are clearly shown in the Voronoi construction in Fig.~\ref{assembly}(3C). When the grafting density is further increased to 0.16$\sigma^{-2}$ and 0.24$\sigma^{-2}$, as in Figs.~\ref{assembly}(4A-4C) and Figs.~\ref{assembly}(5A-5C), respectively, the situation is similar with the NPs forming a single hexagonal layer on top of the brush after the solvent is evaporated. The lattice usually still contains defects which can move around or disappear over time. However, the removal of a defect typically requires long time, indicating that defects can be quite persistent. 

Systems with $A_{\rm np} = 200\epsilon$ behave differently before and during solvent evaporation. For $A_{\rm np} = 200\epsilon$ the NP-brush interaction is strong enough to compete with the entropic penalty associated with embedding a NP in the brush. In the equilibrium state before evaporation, the number of NPs engulfed in the polymer brush grows as $\phi$ increases for the range of $\phi$ studied here and is expected to decrease as $\phi$ increases further, reflecting a balance between the enthalpic gain from NP-polymer attractions and the entropic penalty of embedding NPs in the brush. At a low grafting density $\phi = 0.01\sigma^{-2}$, the NPs are adsorbed onto the domains formed by polymer chains and tend to penetrate the regions with a thickness comparable to the NP diameter after solvent evaporation. In the regions where the brush thickness is larger than the NP diameter, some NPs are embedded in the brush and can even form more than one layers, as shown in Fig.~\ref{assembly}(1D). When the grafting density is quadrupled to 0.04$\sigma^{-2}$, the coverage of brush on the substrate increases but holes still form where there are neither polymer chains nor NPs. In the area coated by the brush, most NPs are engulfed by polymer chains and form a layer close to the substrate [Fig.~\ref{assembly}(2D)]. Since not all the NPs can be accommodated by the brush which only partially covers the substrate, some NPs are adsorbed on top of the first layer and usually stay in registry with the first layer (i.e., the projection of those NPs falls between the NPs in the first layer). The Voronoi constructions shown in Figs.~\ref{assembly}(1F) and (2F) are for the first layer of NPs. Although the overall packing quality is low, some local hexagonal order can still be observed.

The situation is quite interesting at a grafting density $\phi = 0.08\sigma^{-2}$, where the brush is dense enough to fully cover the substrate and its thickness is similar to the NP diameter [Fig.~\ref{assembly}(3D)]. In this case the NPs are all fully embedded in the brush after the evaporation of the solvent and form a single layer with almost perfect hexagonal order as shown in Figs.~\ref{assembly}(3E-3F). Matching the brush thickness and the NP coverage is important for obtaining the well ordered monolayer as can be seen in the data for thicker brushes. When the grafting density is further increased, the polymer chains in the brush are more stretched because of the steric repulsion between the chains, which overcomes the entropy penalty of chain stretching. As a result, the brush thickness is increased to more than twice the NP diameter, which means that the brush is able to engulf more than one layers of NPs. Figs.~\ref{assembly}(4D-4F) and Figs.~\ref{assembly}(5D-5F) show results for $\phi = 0.16\sigma^{-2}$ and $0.24\sigma^{-2}$, respectively. At $C=0.65$, some NPs penetrate deep into the brush while more NPs are embedded in a region close to the brush surface. These latter NPs form a layer in which hexagonal packing is frequently seen but defects are abundant, as shown in the Voronoi constructions for the top layers in Figs.~\ref{assembly}(4F) and (5F). The packing order in this layer is much worse than that in the system with $\phi = 0.08\sigma^{-2}$ [Fig.~\ref{assembly}(3F)]. If the NP concentration in the solution before evaporation is increased to $C \gg 1$, then a 3-dimensional lattice is expected for NPs embedded in a brush with a thickness much larger than the NP diameter. Such a scenario is interesting for future studies.

Solvent evaporation is intrinsically an out-of-equilibrium process and as a result the final distribution of NPs in the polymer brush observed in our simulations may not be an equilibrated one. In particular, our data show that in the final stage of drying, NPs diffuse very slowly inside the brush for strong NP-brush interactions and dense brushes. Instead, only local vibrations of NPs around their average locations are observed in simulations. The results in Figs.~\ref{assembly}(4D-4F) and Figs.~\ref{assembly}(5D-5F) may thus actually correspond to kinetically trapped states. However, these states do not seem to evolve much on time scales accessible to MD simulations and even in additional annealing runs.

\begin{figure}[ht]
\centering
\includegraphics[width=3in]{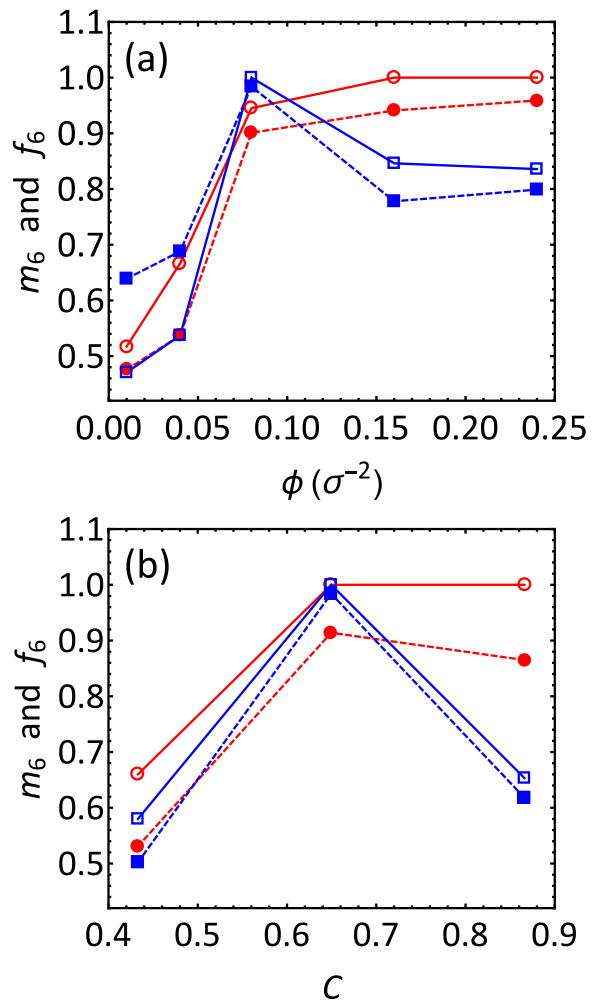}
\caption{The fraction of NPs having 6 neighbors, $f_6$ (open symbols), and the Nelson-Halperin order parameter, $m_6$ (solid symbols), (a) as a function of $\phi$ when $C$ is fixed at 0.65, and (b) as a function of $C$ when $\phi$ is fixed at $0.08\sigma^{-2}$. The data are for the systems with $A_{\rm np} = 100\epsilon$ (circles) and $A_{\rm np} = 200\epsilon$ (squares), respectively. Lines are guides to the eye.}
\label{bond_order}
\end{figure}

To quantify the packing order, i.e., how close a lattice is to an ideal hexagonal lattice, we have computed two parameters. One is $f_6$, the fraction of NPs having exactly 6 neighbors. Another is the Nelson-Halperin order parameter for bond orientation defined as\cite{nelson79}
\begin{equation}
m_6 =  \left| \frac{1}{N_l}\sum_{j=1}^{N_l}
\frac{1}{n_j}\sum_{k=1}^{n_j}\exp (i6\theta_{jk})\right| ,
\label{NHEq}
\end{equation}
where the first sum is over all NPs in the lattice, the second sum is over all the $n_j$ nearest neighbors of the $j$-th NP, and $\theta_{jk}$ is the angle formed by the bond between the $j$-th NP and its $k$-th nearest neighbor with respect to a fixed axis. From the definition $f_6=1$ and $m_6=1$ indicate an ideal hexagonal lattice and smaller values correspond to lattices with disorder and defects. Results for $f_6$ and $m_6$ for the systems in Fig.~\ref{assembly} are shown in Fig.~\ref{bond_order}(a). The two parameters follow the same trends as $\phi$ varies. In particular, for $A_{\rm np} = 200\epsilon$ both $f_6$ and $m_6$ peak at $\phi=0.08\sigma^{-2}$, indicating that a polymer brush with an optimized grafting density is required to produce a well-packed 2D hexagonal lattice of NPs [see Fig.~\ref{assembly}(3D)] when the NP-brush interaction is strong. For $A_{\rm np} = 100\epsilon$, however, the value of $f_6$ and $m_6$ are close to 1 as long as $\phi\gtrsim 0.08\sigma^{-2}$, indicating that a dense brush is capable to adsorb a layer of NPs on its surface and the NPs naturally order into a hexagonal lattice with few defects.

\begin{figure}[ht]
\centering
\includegraphics[width=3.25in]{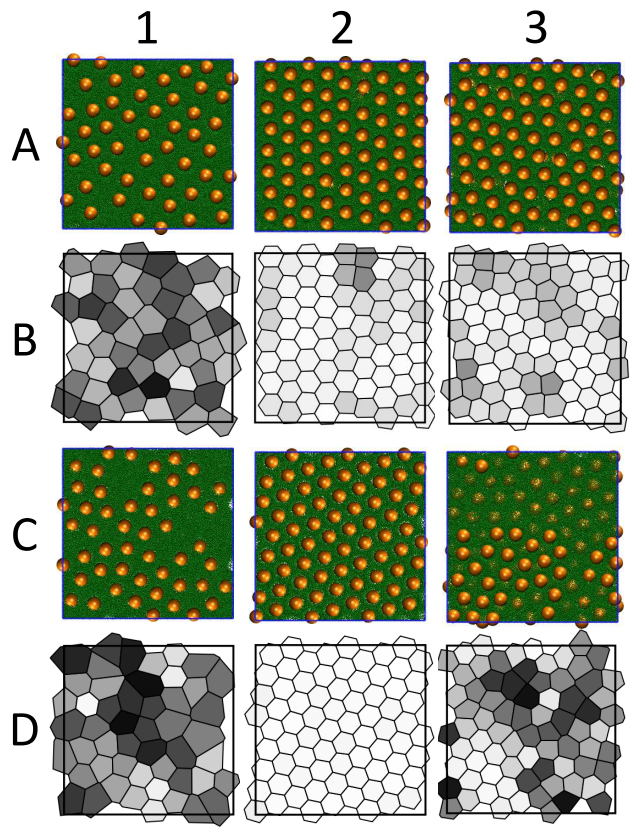}
\caption{The organization of NPs at various coverages on top of a brush with $\phi=0.08\sigma^{-2}$ after solvent evaporation. (A) and (C) top view; (B) and (D) Voronoi construction. (1) $C=0.43$, (2) $C=0.65$, and (3) $C=0.87$. For the three systems in (A) and (B), $A_{\rm np} = 100\epsilon$. For (C) and (D), $A_{\rm np} = 200\epsilon$. Note that in (1C)-(3C) all NPs are embedded in the brush; for better visibility a layer of polymer beads at the top is removed. Note that (2C) and (2D) here are identical to Fig.~\ref{assembly}(3E) and Fig.~\ref{assembly}(3F), respectively. They are duplicated to make comparison more convenient. The color codes are the same as in Fig.~\ref{assembly}.}
\label{packing}
\end{figure}

In Fig.~\ref{packing} the effect of NP coverage on the resulting lattice of NPs after solvent evaporation is shown for $\phi=0.08\sigma^{-2}$. The top two rows are for $A_{\rm np} = 100\epsilon$. At a low NP coverage ($C=0.43$), the packing is random and no hexagonal order is observed as shown in Fig.~\ref{packing}(1B). When the NP coverage is increased to $C=0.65$, a hexagonal lattice emerges after the evaporation of the solvent. Note that the results in Figs.~\ref{packing}(2A-2B) are obtained in a second, independent run from the one used to generate Figs.~\ref{assembly}(3A-3C). The fact that in all of our simulations the partially adsorbed NPs order into a hexagonal packing on top of the brush after evaporating the solvent shows that the result is quite robust. When the NP coverage is further increased to $C=0.87$, some NPs diffuse into the brush but the rest form a hexagonal lattice with few defects at the surface of the brush, as shown in Figs.~\ref{packing}(3A-3B), confirming that the hexagonal order occurs at the brush surface as long as the NP coverage is high enough and the brush is dense enough to yield a relatively uniform thickness. The results on $f_6$ and $m_6$ in Fig.~\ref{bond_order}(b) show the same trend.

\begin{figure*}[ht]
\centering
\includegraphics[width=6in]{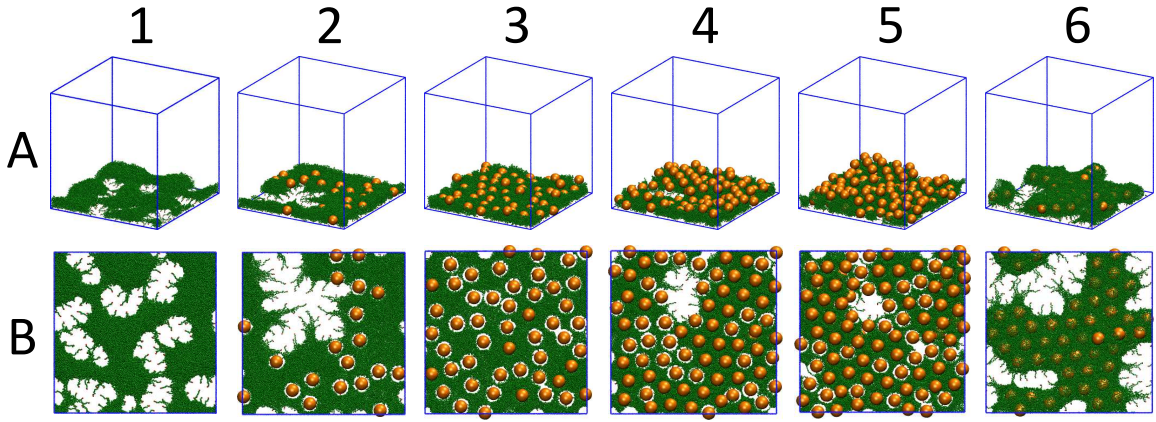}
\caption{The adsorption of NPs at various coverages in a brush with $\phi=0.04\sigma^{-2}$ after solvent evaporation. (A) side view and (B) top view. (1) dry brush, (2) $C=0.17$, (3) $C=0.43$, (4) $C=0.65$, (5) $C=0.87$, and (6) $C=0.43$. The NP-polymer interaction strength is $A_{\rm np} = 100\epsilon$ for (2)-(5) while $A_{\rm np} = 200\epsilon$ for (6). The color codes are the same as in Fig.~\ref{assembly}.}
\label{healing}
\end{figure*}

In Fig.~\ref{bond_order}(b) and the bottom two rows of Fig.~\ref{packing}, the results for the systems with $\phi=0.08\sigma^{-2}$ and $A_{\rm np} = 200\epsilon$ are also included. For $C=0.43$, all the NPs are embedded at similar heights in the brush, forming a single layer. However, the NP coverage is low and only local hexagonal packing occurs within disorder at large scales as shown in Fig.~\ref{packing}(1C). For $C=0.65$, a hexagonal NP lattice with few if any defects is formed. For $C=0.87$, about 80\% of the NPs penetrate into the brush and form a layer close to the substrate. The Voronoi construction for this first layer is shown in Fig.~\ref{packing}(3D). The remaining NPs are randomly distributed above the first layer. Therefore, for $A_{\rm np} = 200\epsilon$ a 2D hexagonal lattice of NPs is only obtained when both $\phi$ and $C$ are fine tuned to yield a system where a single layer of NPs with high enough loading is embedded in the brush. The same trend is also shown in the quantitative results in Fig.~\ref{bond_order} for $f_6$ and $m_6$ as functions of $\phi$ and $C$, respectively. A brush with a lower $\phi$ than the optimal value cannot fully cover the whole substrate [e.g., Figs.~\ref{assembly}(2D-2F)] while at a higher $\phi$ the brush is thick enough to support two layers of NPs [e.g., Figs.~\ref{assembly}(5D-5F)] or more. A system with $C$ lower than optimal has NPs forming a single layer in the brush but the layer is too dilute for hexagonal packing [e.g., Figs.~\ref{packing}(1C-1D)]. For systems with $C$ higher than optimal, NPs form more than one layers with diminished or no hexagonal order in each layer [e.g., Figs.~\ref{packing}(3C-3D)]. It should be pointed out that when $C$ is higher than optimal, the distribution of NPs engulfed in the brush varies with $\phi$. In Figs.~\ref{packing}(3C-3D) for $\phi=0.08\sigma^{-2}$ and $C=0.87$, the majority of the NPs first forms a layer close to the substrate with the rest distributed above this layer. However, when $\phi$ is increased to $0.16\sigma^{-2}$ and $0.24\sigma^{-2}$, most of the NPs engulfed in the brush form a layer close to the brush surface with the rest distributed below this layer, as shown in Figs.~\ref{assembly}(4D-4F) and Figs.~\ref{assembly}(5D-5F) for $C=0.65$. The same phenomenon is also observed for $\phi=0.24\sigma^{-2}$ and $C=0.87$ with a major layer of NPs near the brush surface (results not shown). This change is consistent with the idea that there is a larger entropic penalty of penetrating a denser polymer brush.

Finally we have observed an interesting effect brought about by the adsorption of NPs on a polymer brush with a low grafting density. The results in Fig.~\ref{healing} are for a brush with $\phi = 0.04\sigma^{-2}$ at various values of NP coverage. For the dry brush (i.e., $C=0$), the chains are not sufficient to cover the whole substrate and form domains that percolate the plane of the substrate as shown in Figs.~\ref{healing}(1A-1B); many holes can be seen. At $C=0.17$, the brush covers most of the surface with only one hole. When the NPs are adsorbed on top of the brush for the case of a weak NP-polymer interaction (e.g., $A_{\rm np} = 100\epsilon$), the thickness becomes more uniform. When the NP coverage is increased to $C=0.43$, the hole completely disappears and the brush is able to fully cover the substrate, which we call a ``healing'' effect of the adsorbed NPs. Interestingly, the hole reappears when the NP coverage is increased to larger values [Figs.~\ref{healing}(4A-4B) and (5A-5B)], indicating the more NPs do not strengthen the ``healing'' effect, though the brush is partially ``healed'' if we compare Figs.~\ref{healing}(4B) and (5B) to Fig.~\ref{healing}(1B). One possible issue is that the high evaporation rates resulting from evaporating the solvent into a vacuum may play a role in the ``healing'' capability of partially adsorbed NPs. Within our limited range of evaporation rates, we find the same behavior, but much slower rates would be an interesting direction for future studies.

The ``healing'' effect was never observed for strong NP-polymer interactions. For example, for $A_{\rm np} = 200\epsilon$ ($\theta_c = 0$) the brush is still highly nonuniform for $C=0.43$ [Figs.~\ref{healing}(6A-6B)], which shows a good ``healing'' capability as shown in Figs.~\ref{healing}(3A-3B) for $A_{\rm np} = 100\epsilon$ ($\theta_c = 79^\circ$). These results reveal an interesting strategy to make a low-grafting-density brush more uniform by using adsorbed, but not embedded, NPs at intermediate values of coverage.

\section{Conclusions}
In this paper we have presented results from large-scale molecular dynamics simulations on ordering NPs with a polymer brush via solvent evaporation. It has been found that a dense brush with grafting density $\phi \gtrsim 0.08\sigma^{-2}$ and a weak NP-polymer interaction resulting in NP adsorption on top of the brush can be combined to robustly yield a hexagonally packed lattice of NPs after solvent evaporation as long as the NP coverage $C$ is in the range leading to a single layer that is not too dilute. If the NP-polymer interactions are strong, then a system with an appropriate combination of $\phi$ and $C$ is required to yield a single layer of NPs engulfed in the brush with a thickness similar to the NP diameter. This layer of NPs has a well-ordered hexagonal arrangement. Our results thus point to strategies to generate a 2D hexagonal lattice of NPs using a polymer brush depending on the interactions between the NPs and the brush. Our results also reveal an interesting effect that the adsorption of NPs can ``heal'' a polymer brush with a low grafting density and make the brush to be more uniform and even fully cover the substrate. This discovery may lead to interesting applications of brush-NP hybrid systems as a protective layer of a solid surface.

\section*{Acknowledgments}
Acknowledgment is made to the Donors of the American Chemical Society Petroleum Research Fund (PRF \#56103-DNI6), for support of this research. This research used resources of the National Energy Research Scientific Computing Center (NERSC), which is supported by the Office of Science of the United States Department of Energy under Contract No. DE-AC02-05CH11231. This work was performed, in part, at the Center for Integrated Nanotechnologies, an Office of Science User Facility operated for the U.S. Department of Energy (DOE) Office of Science. Sandia National Laboratories is a multimission laboratory managed and operated by National Technology and Engineering Solutions of Sandia, LLC., a wholly owned subsidiary of Honeywell International, Inc., for the U.S. Department of Energy's National Nuclear Security Administration under contract DE-NA-0003525.

%\bibliography{allref.np}

\end{document}